\documentclass[10pt,A4paper,twocolumn,aps,pra, superscriptaddress,longbibliography, nofootinbib]{revtex4-2}

\usepackage{graphicx}
\usepackage{dcolumn}
\usepackage{bm}
\usepackage{gensymb}
\usepackage{float}

\begin{document}

\preprint{AIP/123-QED}

\title{Fabrication and characterization of low-loss Al/Si/Al parallel plate capacitors for superconducting quantum information applications}


\author{Anthony P. McFadden}
\email{anthony.mcfadden@nist.gov}
\affiliation{National Institute of Standards and Technology, Boulder CO, 80305}

\author{Aranya Goswami}%
\altaffiliation[Now at ]{MIT}
\affiliation{ 
Department of Electrical and Computer Engineering, University of California, Santa Barbara 93106
}

\author{Tongyu Zhao}%
\affiliation{ 
Physics Department, University of Colorado, Boulder, 80302
}
\affiliation{National Institute of Standards and Technology, Boulder CO, 80305}

\author{Teun van Schijndel}
\affiliation{ 
Department of Electrical and Computer Engineering, University of California, Santa Barbara 93106
}

\author{Trevyn F.Q. Larson}
\author{Sudhir Sahu}
\author{Stephen Gill}
\author{Florent Lecocq}
\author{Raymond Simmonds}
\affiliation{National Institute of Standards and Technology, Boulder CO, 80305}

\author{Chris Palmstr\o m}
\affiliation{ 
Department of Electrical and Computer Engineering, University of California, Santa Barbara 93106
}
\affiliation{ 
Materials Department, University of California, Santa Barbara 93106
}

\date{\today}

\begin{abstract}
Increasing the density of superconducting circuits requires compact components, however, superconductor-based capacitors typically perform worse as dimensions are reduced due to loss at surfaces and interfaces. Here, parallel plate capacitors composed of aluminum-contacted, crystalline silicon fins are shown to be a promising technology for use in superconducting circuits by evaluating the performance of lumped element resonators and transmon qubits. High aspect ratio Si-fin capacitors having widths below $300nm$ with an approximate total height of 3$\mu$m are fabricated using anisotropic wet etching of Si(110) substrates followed by aluminum metallization. The single-crystal Si capacitors are incorporated in lumped element resonators and transmons by shunting them with lithographically patterned aluminum inductors and conventional $Al/AlO_x/Al$ Josephson junctions respectively. Microwave characterization of these devices suggests state-of-the-art performance for superconducting parallel plate capacitors with low power internal quality factor of lumped element resonators greater than 500k and qubit $T_1$ times greater than 25$\mu$s. These results suggest that Si-Fins are a promising technology for applications that require low loss, compact, superconductor-based capacitors with minimal stray capacitance.

\end{abstract}

\maketitle

\section{\label{sec:level1}Introduction}

Superconducting qubits employ Josephson junctions, inductors, capacitors, and transmission lines to realize physical systems exhibiting quantum properties with discrete energy levels that can be used for computation\cite{Nakamura1999,Koch2007,Manucharyan2009}. Quantum computation using superconducting qubits has been shown to be possible through milestone demonstrations moving toward quantum advantage\cite{Arute2019,Kim2023}. Studies demonstrating large scale qubit integration\cite{Chow2021} have shown that superconducting qubit technology is promising, and is presently accepted as a leading candidate to achieve large-scale, fault tolerant quantum computation\cite{Kjaergaard2020}.

The  Josephson junction and the reactive components of qubit circuits are made from superconductors and insulators that are dissipationless in principle; however, microscopic contamination, such as undesired oxides and residues at surfaces and interfaces, can serve as dissipation channels which cause decoherence in real systems\cite{deLeon2021,Siddiqi2021}. These oxides and residues are known to host two-level systems (TLS)\cite{Simmonds2004}, which may exchange energy with qubits and are understood to be the dominant decoherence source in many real devices\cite{Shalibo2010,Lisenfeld2019,McRae2020,McRae2020b}.

Coherence times in transmons, one type of superconducting qubit which consists of a Josephson junction shunted by a capacitor, have dramatically improved in recent years due to the integration of new materials and mitigation of surface loss\cite{Place2021,Wang2022,Crowley2023}. Mitigation of surface loss through design of the transmon geometry has proven effective, where large, planar capacitor structures are often used to dilute the participation of the surfaces and interfaces. These large (typically millimeter scale) devices tend to outperform more compact capacitor structures such as parallel-plate and inter-digitated designs\cite{Wang2015, Gambetta2017}. However, a major disadvantage of larger planar capacitors is a reduction in scalability, both due to the increased footprint on chip and the need to address stray fields that couple devices to other resonant modes on a chip, as well as the packaging environment. In addition, these large devices have electric field energy residing at surfaces that are exposed to the ambient atmosphere and contamination sources during processing, storage, and packaging steps, which brings consistency, repeatability, and age resilience into question.

Many of these issues may be addressed by switching to parallel plate capacitor (PPC) geometries composed of superconductor/insulator/superconductor (SIS) sandwiches. PPCs offer advantages over planar capacitor geometries\cite{Zhao2020,Mamin2021}: the electric field energy is mostly contained in the tri-layer structure with little participation in surface regions exposed to atmosphere, and are also inherently more compact. However, having the field energy contained within the dielectric region of the sandwich places a high demand on the loss properties of the dielectric and its interfaces with the superconductor. 

Fabrication of PPCs and Josephson junctions from SIS tri-layers with deposited dielectrics or vacuum gap capacitors has been demonstrated\cite{Oh2006,Cicak2010,McFadden2020,Weber2011,Patel2013}; however devices made using tri-layer processes typically have high microwave loss and are not used for applications such as transmons with high coherence times. Conventional growth of planar SIS tri-layers is difficult, largely due to materials considerations in growing a high quality dielectric on a superconducting metal. Interestingly, at the time of this writing, the best performing PPC from the standpoint of low-loss microwave performance was made by exfoliating 2-D materials rather than direct tri-layer growth\cite{Antony2021,Wang2022a}.

In this work, an alternative approach is used to create parallel plate capacitors by etching the capacitor dielectric out of Si substrates\cite{Goswami2022}, which has been established as one of the lowest loss dielectrics available\cite{Checchin2022}. The process presented in this work takes advantage of the highly anisotropic crystallographic etching of Si(110) substrates with potassium hydroxide (KOH) to form high-aspect-ratio fin structures with flat Si\{111\} sidewalls which are contacted with aluminum to form the PPC. These Si fins have been incorporated as the capacitive element in both lumped element resonators and transmons. We find that the Si fins are high performing with figures of merit including low power internal quality factor $(Q_i)$ of lumped element resonators, qubit quality factor, and qubit $T_1$ times meeting or exceeding the state-of-the-art for comparable devices made using other parallel plate capacitor technology.

\section{Methods}

\subsection{Lumped element resonator and transmon design}

Frequency multiplexed lumped element resonator chips and transmon chips with multiplexed readout resonators were fabricated on the same $50.8mm$ wafer. Having both resonator and transmon chips fabricated together allows for independent characterization of the Si fin PPC and the transmon. 3D simulations of the resonators and qubits were performed using finite element analysis.

The lumped element chip design consists of 8 resonators inductively coupled to a central feedline that are measured in the hanger configuration. Each lumped$-$element resonator consists of a metallized section of a Si fin (capacitor), which is shunted by a $15\mu m$ wide Al thin-film wire inductor having an inductance of $4.29nH$ and stray capacitance of $84.1fF$ determined by simulation. For the devices presented here, $81-84\%$ of the total capacitance of each lumped element resonator is from the fin capacitor. The frequency of the resonators is controlled by lithographically incrementing the metallized length along the fin capacitor, which is linearly related to the device capacitance. 

The transmon chip layout consists of 6 nominally identical transmon qubits coupled to frequency multiplexed readout resonators. Devices were designed to employ Si fin parallel plate capacitors shunted by Josephson junctions. The transmons were designed to have anharmonicities $\alpha/2\pi=300 MHz$, qubit-readout coupling constants $g/2\pi=100 MHz$, and $E_J/E_C = 50$.  

The readout resonators are quarter$-$wave coplanar waveguide (CPW) resonators inductively coupled to a central feedline and capacitively coupled to the fin transmons. In this design, both the coupling capacitance and the transmon capacitance are realized using metallized sections of the same Si fin, demonstrating the convenience of the Si-fin process to construct capacitive elements.

\begin{figure*}[hbt!]
\includegraphics[width=1.0\textwidth]{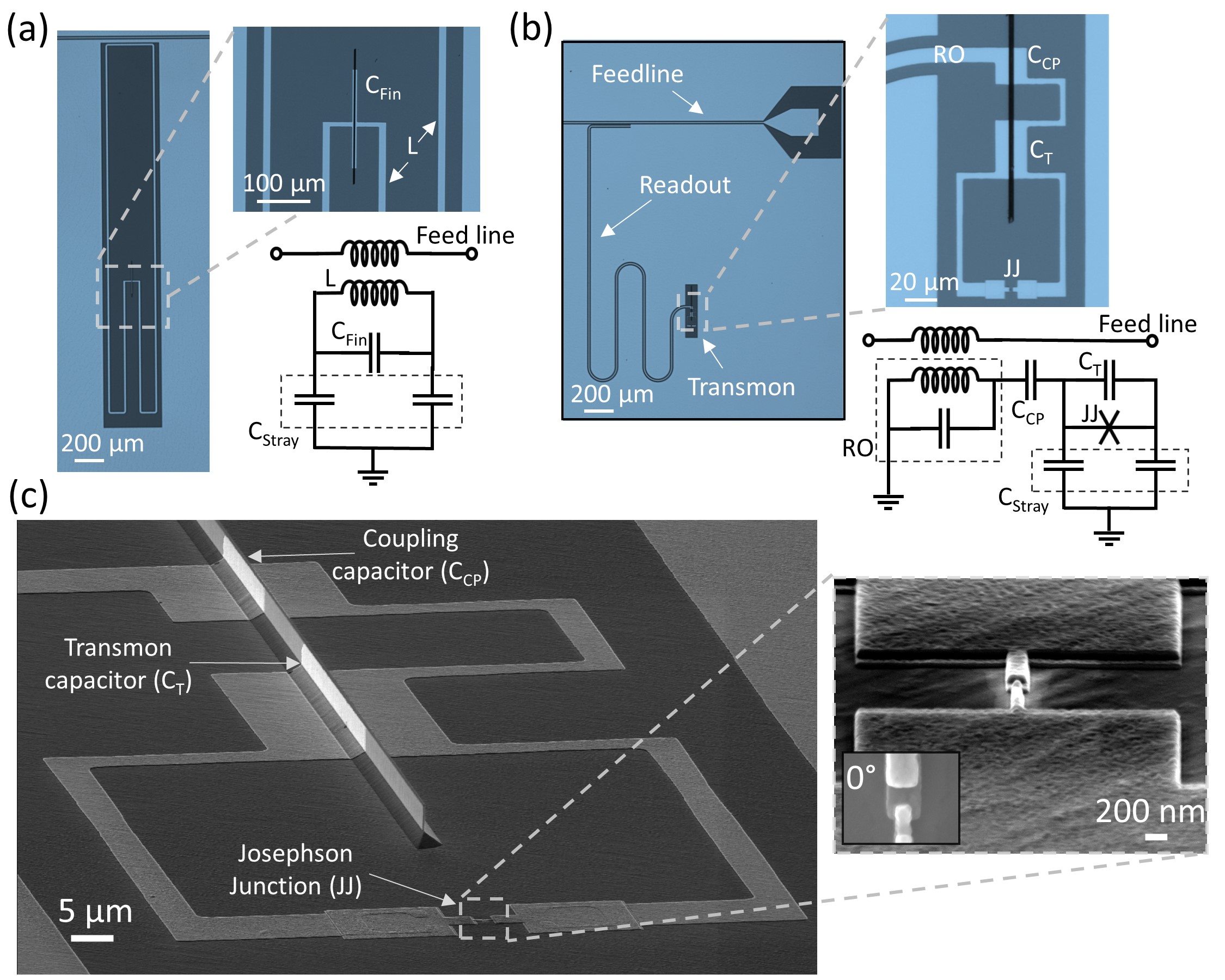}
\caption{\label{SEM} (a), (b) Optical micrographs and corresponding circuit schematics for Si-fin lumped element resonators and transmons respectively. The functional components of the fin transmon are indicated and illustrate the capacitive coupling of the fin transmon to the readout resonator (RO). (c) Scanning electron micrograph of a Si-fin transmon. The insets show the Josephson junction region in higher magnification.}
\end{figure*}

\subsection{Device fabrication}

Fin capacitors are fabricated on Si(110) substrates using a combination of dry and wet etching closely following the process outlined in a previous work \cite{Goswami2022}. As-received substrates were cleaned before LPCVD growth of 100nm low-stress silicon nitride $(SiN_x)$ that serves as a hard mask. The fins are patterned using electron beam lithography(EBL) and a negative resist followed by dry etching to pattern the LPCVD nitride. The fin patterns are aligned so that the long axis of the fin is parallel to the major flat of the Si(110) wafer, which is critical to obtain the desired flat $\{111\}$ sidewalls and high aspect ratio structures.

After resist stripping and wafer cleaning, the wafer was immersed in a solution of 45$\%$ KOH (by weight) in water held at $87\degree C$ for 2 minutes to anisotropically etch the Si. This procedure results in approximately $3\mu m$ etching into the substrate with approximately 40nm of undercut on each side of the $SiN_x$ hard mask. The resulting fin structure consists of a $2.2\mu m$ tall parallel-plate region having uniform thickness near $220nm$ with flat $Si\{ 111 \}$ sidewalls and a tapered base region having an approximate height $0.8\mu m$.

Following formation of the Si-Fin capacitors, the wafer was cleaned with solvents, and the silicon native oxide was removed with a buffered oxide etch (BOE). After the BOE etch, the wafer was rinsed in DI water before immediately loading in high vacuum for aluminum deposition. Aluminum was deposited in two $50nm$ thick depositions at $\pm 25\degree$ from the wafer surface normal and aligned to coat the sides of the fins. This procedure results in $100nm$ of aluminum metal deposited on the wafer surface and $50nm$ on the sides of the fins. The undercutting of the fins below the $(SiN_x)$ hard mask was used to shadow the aluminum during deposition to form a self aligned break in the metallization near the top of the fins.

Following aluminum deposition, the metal layer was patterned using optical lithography and wet etching. A standard spin-on photoresist was used, using a spin process that results in a layer thickness of approximately 1$\mu m$ on an unpatterned wafer. On wafers with patterned fins, the resist profile is nonuniform near the fins due to the topography. In order to accurately transfer patterns onto the wafer, extra exposure doses were applied near the fins as needed to clear resist residue. After photolithography, the $100nm$ thick Al layer was wet-etched using a phosphoric and nitric acid based aluminum etchant at room temperature.

$Al/AlO_x/Al$ Josephson junctions were deposited using a standard EBL/Dolan bridge process\cite{Dolan1977} employing a PMMA/PMGI bi-layer. The target dimensions of the junctions was $100nm$ x $100nm$. As previously stated, resist profiles are nonuniform in proximity to the fin capacitors, so junction dimensions were calibrated on a separate wafer by patterning junctions at varying distances from the ends of the fins and measuring dimensions with atomic force microscopy. It was found that placing junctions at distances greater than $30\mu m$ from the ends of the fin resulted in reproducible junction dimensions within $10\%$ of the targeted values.

Following EBL exposure and development, the wafer was loaded in an electron beam evaporator for Josephson junction deposition. Prior to bottom electrode deposition, the exposed areas of the wafer were ion milled to remove native oxide from the Al metallization layer and create transparent contacts to the junction. A $35nm$ thick bottom electrode was deposited at $+ 14.7\degree$, oxidized at $180mTorr$ in pure $O_2$ for 10 minutes to form the tunneling barrier, followed by deposition of the $75nm$ thick bottom electrode at $- 14.7\degree$. After junction deposition, protective photoresist was spun on the wafer which was diced down to $7.5mm$ x $9.5mm$ die. Individual chips were then selected and the aluminum layer from junction deposition was lifted off. The finished junctions measured after dicing had room temperature resistances near $9.3k\Omega$ suggesting an approximate junction critical current of $30nA$ and Josephson inductance of $11nH$. Figure \ref{SEM} shows optical micrographs and circuit schematics for both the lumped element resonator and transmon devices as well as a scanning electron micrograph of a fabricated transmon.

\subsection{Measurement Setup}

After lift-off, cleaned chips were wire-bonded with aluminum wire and packaged in a 2-port package constructed of Au-plated OFHC Cu for low temperature RF measurements. All microwave measurements were performed in a dilution refrigerator having a base temperature of 35 mK. The attenuation and amplification setup used for both resonators and qubits is similar to that presented by McRae et al.\cite{McRae2020b}. Lumped element resonators were characterized using a vector network analyzer (VNA) by fitting the power dependent $S_{21}$ spectra of the resonators using the diameter correction method (DCM)\cite{Khalil2012,McRae2020b}. Qubits were measured in time domain using QICK software running on a Xilinx/AMD ZCU216 RF SoC board with a measurement configuration similar to that presented by Stefanazzi et al.\cite{Stefanazzi2022}.

\section{Results and Discussion}

\begin{figure*}[hbt!]
\includegraphics[width=1.0\textwidth]{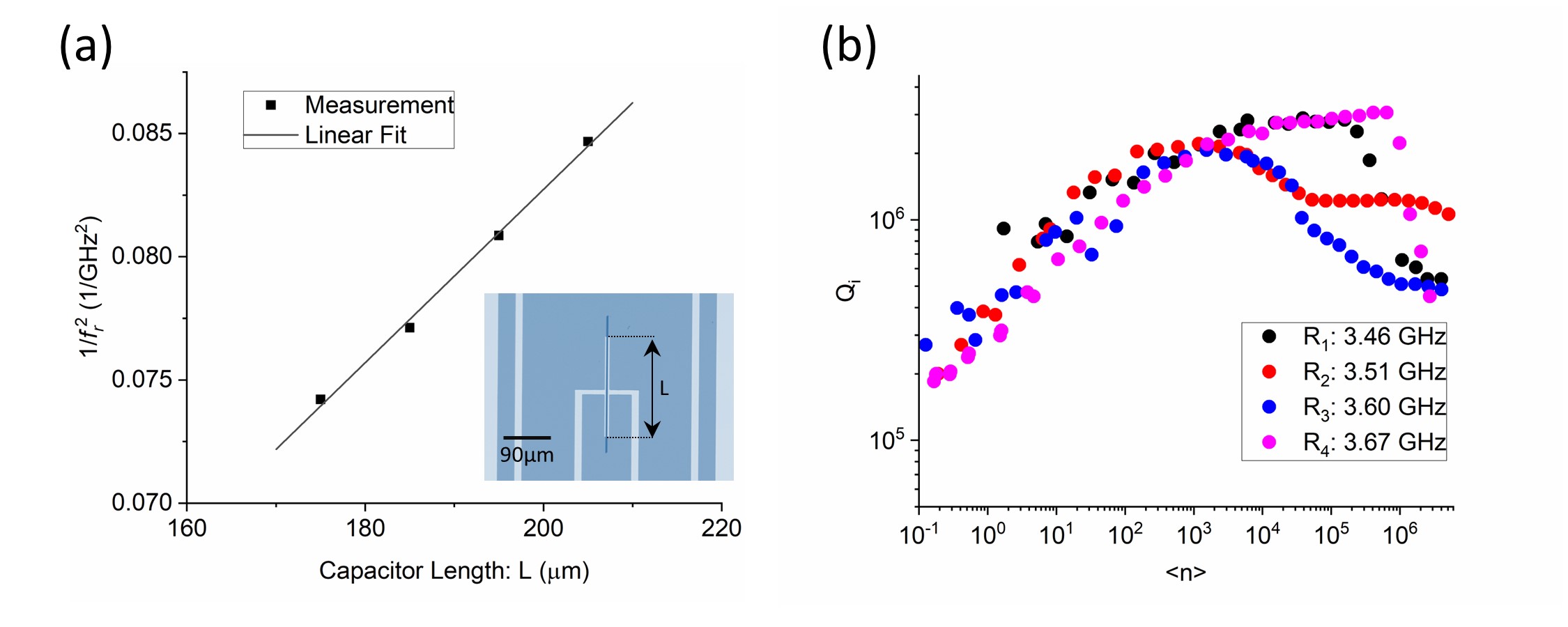} 
\caption{\label{LERTransport} Measurement results from RF characterization of lumped element resonators (LER). (a) Measured resonant frequency vs metallized capacitor length (L) for LERs. The inset includes an optical micrograph of a device with the dimension L indicated. Electromagnetic simulations of the inductor loop suggest that the fin structure contributes 81-84\% of the total resonator capacitance. (b) Extracted internal quality factor ($Q_i$) plotted against estimated photon number in the cavity of the four resonators measured in (a).}
\end{figure*}

Figure \ref{LERTransport}(a) shows the measured resonant frequency of the lumped element resonators as a function of metallized length along the Si fins. The data suggests the expected linear dependence of $(1/f_r)^2$ on capacitor length. Given the inductance value of $4.29nH$ determined from simulations, the slope of the interpolated line may be used to estimate the capacitance per unit length $(C_0)$ of the fin structures which is found to be $2.1fF/ \mu m$. Extrapolating the linear fit to zero fin length ($L=0$) gives a resonant frequency of $8.97 GHz$, in reasonable agreement with simulation results yielding a self-resonance of the inductor loop at $8.3 GHz$.

A crude parallel plate approximation of fin capacitance using the measured fin height and width of $2.2\mu m$ and $220nm$, respectively, yields $\approx1.0fF/\mu m$ for a silicon dielectric constant of 11.7. This significantly lower value reflects the contribution of stray and additional capacitance near the base of the fin to the total capacitance of the structures. Notably, the electric fields from metallization at the base of the fin are predominantly contained in the Si substrate rather than at the material surfaces.

Power dependent $S_{21}$ of the lumped element resonators was also measured and is included as Fig \ref{LERTransport}(b). Internal and coupling quality factors ($Q_i$  and  $Q_c$) were extracted from the $S_{21}$ data, and the cavity photon number was then estimated using $Q_i$, $Q_c$, and the measured input line attenuation.

Referring to Fig. \ref{LERTransport}(b), the lumped element resonator $Q_i$ reaches a maximum at intermediate powers and anomalously decreases with increasing power. This effect is attributed to Cooper-pair breaking in the superconducting aluminum\cite{deVisser2014,McRae2020b} at higher RF drive power and has been reported before in similar structures\cite{Wang2022a}. $Q_i$ values near the single photon level are observed to be very good, with several points exceeding $500k$, which, to the authors’ knowledge, is the current state-of-the-art for superconducting resonators made using parallel plate capacitors\cite{Wang2022a}.

\begin{figure}[hbt!]
\includegraphics[width=0.5\textwidth]{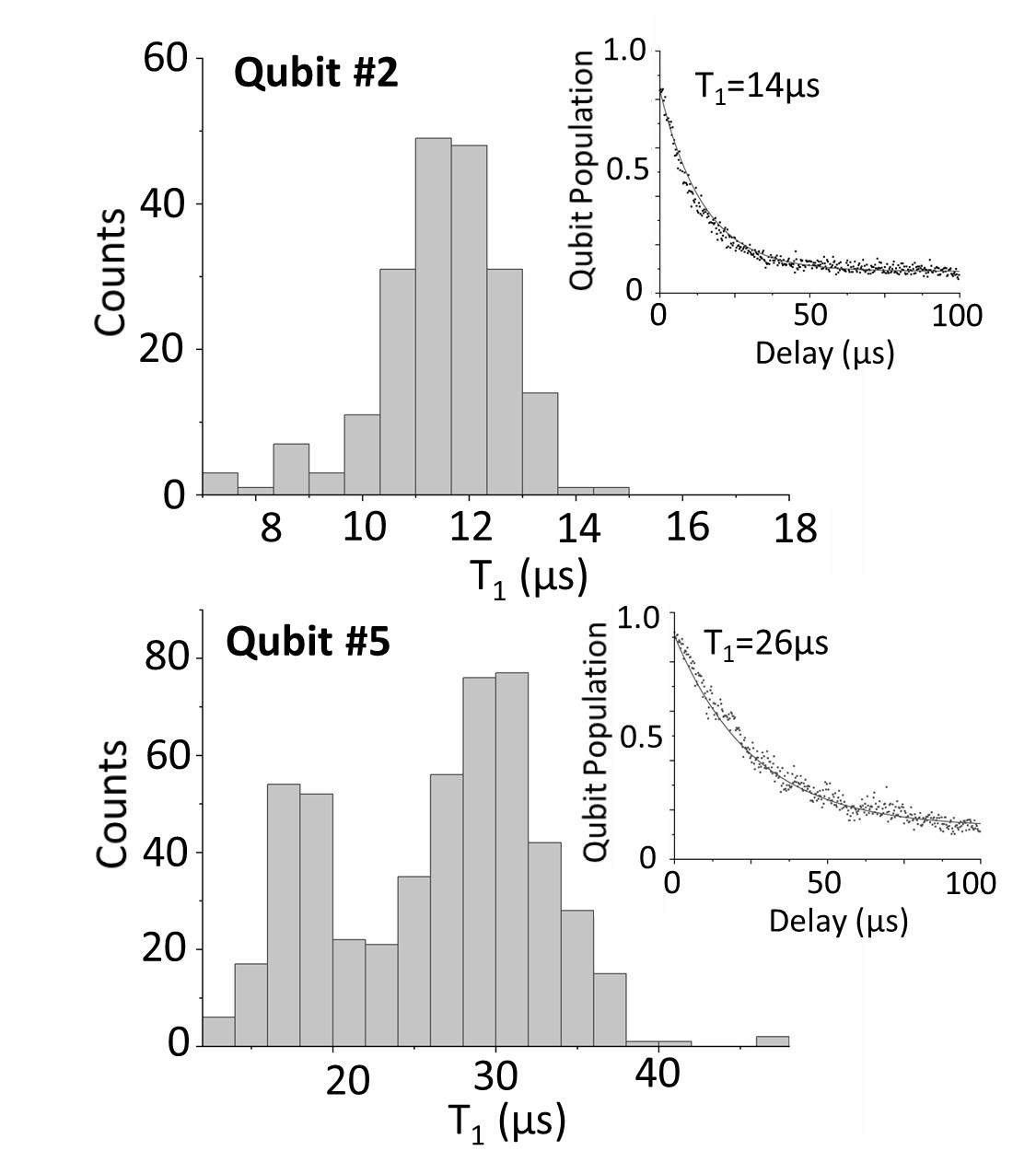} 
\caption{\label{transmonMeas} Histograms of measured $T_1$ times for qubits \#2 and \#5. The histogram for qubit \#5 appears bimodal, likely due to the intermittent interaction of the qubit with two level systems. The insets show example $T_1$ traces along with exponential fits.}
\end{figure}

All 6 qubits on the separate transmon chip were found to be functional, and characterization results are included as Table 1. Anharmonicity $(\alpha)$ was computed using the measured values of $f_{01}$ and $f_{02}/2$ obtained from continuous-wave qubit spectroscopy measurements. $f_{01}$ and $f_{02}$ denote the frequencies corresponding to the $0\rightarrow 1$ and $0\rightarrow 2$ energy transitions respectively. The coupling constant $g$ was estimated by measuring the frequency shift of the readout resonator between the bare and coupled states (punch-out)\cite{Reed2010}, using $\omega_{coupled}-\omega_{bare}=g^2/\Delta$, where $\Delta$ is the detuning between the qubit and resonator. The coupling constant $g$ may also be calculated using the measured dispersive shift from the $0\rightarrow 1$ transition and employing the approximation $\chi \approx g^2/\Delta-g^2/(\Delta-\alpha)$. We obtain similar but slightly higher values of $g$ using the dispersive shift approximation with a somewhat larger spread between devices. The $E_J$ values used to estimate the $E_J/E_C$ ratio were computed using measured values of the qubit frequencies and anharmonicity. Likely due to a defective junction, the resonance of Qubit $\#1$ was measured to be much higher than the other qubit frequencies, colliding with the readout resonator frequency and preventing extensive characterization.

Transmon anharmonicity, ($\alpha/2\pi$) and qubit to readout coupling constant, ($g/2\pi$) were observed to be quite uniform and near the targeted values, which is notable as these parameters are determined by the fin capacitors. The transmon $T_1$ values were measured to be near $20\mu s$ with mean $T_1$ values for the 6 devices ranging from 11 to 26 $\mu s$. Figure \ref{transmonMeas} shows $T_1$ histograms from qubits \#2 and \#5 taken over the course of several hours. The corresponding qubit quality factors $(Q=\omega T_1)$ are in the range 350k-750k, which are reasonably close to the measured low-power internal quality factors of the lumped element resonators as shown in Fig \ref{LERTransport}(b). Notably, qubit quality factor is expected to be independent of qubit frequency in contrast to $T_1$ time and is taken to be a better figure of merit. Ramsey measurements were also performed, and $T_2^*$ was measured to be near $5\mu s$. The low $T_2^*$ values observed are attributed to photon shot noise in the readout cavity, which is largely determined by the measurement environment and the readout resonator linewidth rather than the transmon capacitor.

\begin{table*}[t]
\caption{\label{qubitData}Measured and computed parameters extracted from Si-fin transmons}
\begin{ruledtabular}
\begin{tabular}{ccccccccccc}
 Qubit\# &$f_{01}$&$f_{RO}$&$\alpha /2\pi $&$E_J/E_C$&$g/2\pi $ &$\kappa /2\pi $ &$T_{Purcell}$ &$T_1 (mean)$&$T_2^* (Ramsey)$&Qubit Q\\
 
  &(GHz)&(GHz)&(MHz)& &(MHz)&(MHz)&($\mu s$)&($\mu s$)&($\mu s$)\\ \hline
 
 1\footnote{Likely defective junction}&$5.5$&$5.675$&-&-&-&-&-&-&-&- \\
 
 2&$4.920$&$5.731$&-270&$46$&98&1.1&10&11&2.6&350k\\
 
 3&$4.547$&$5.758$&-200&$69$&92&0.84&32&15&4.1&430k\\
 
 4&$4.713$&$5.803$&-274&$41$&94&0.90&23&20&4.6&580k\\
 
 5&$4.565$&$5.849$&-279&$38$&93&1.1&28&26&5.6&750k\\

 6&$4.622$&$5.893$&-273&$40$&96&1.2&24&22&4.8&650k\\
\end{tabular}
\end{ruledtabular}
\end{table*}

Purcell time was estimated using the approximation $T_{Purcell}\approx\Delta^2/(g^2\kappa)$, where $\kappa$ is the spectral linewidth of the readout resonator. Referring again to Table 1, the measured $T_1$ values are quite close to, and in some cases even exceed the estimated Purcell limit. This discrepancy is not necessarily nonphysical, as the Purcell rate estimation used here relies on a simplified model that does not include the experimental $S_{21}$ background, modeled admittance of the resonator, and neglects the effects of higher-order resonances in the readout. Including higher order resonances when modeling spontaneous emission has been shown to increase calculated Purcell times when qubits are detuned to frequencies below the readout, as is the case for the measurements presented here\cite{Houck2008}. While the Purcell estimation raises the question of whether the $T_1$ measurements provide an accurate assessment of device performance, it does establish a lower bound on the intrinsic quality factor of the Si-fin transmons, which may be reassessed with new devices following trivial modifications to the readout resonator to increase the Purcell time.
\section{Conclusion}

A fabrication process where parallel-plate capacitors are constructed from low-loss Si substrates has been evaluated for superconducting quantum computing applications. The Si-fin capacitors were incorporated in both lumped element resonators and transmons using established fabrication techniques. RF measurements of the resonators and transmons were performed at 35mK and show that the devices are high performing: meeting or exceeding the state-of-the-art for comparable devices made using parallel-plate capacitors, with low-power $Q_i$ of the best lumped-element resonators exceeding 700k and the best transmon quality factors exceeding 750k. 

The computed Purcell time of the transmons was comparable and in some cases even greater than measured $T_1$ times, suggesting the intrinsic quality factor of these transmons is likely higher. Moreover, the presence of stray capacitance in the LE design suggest that the intrinsic quality factor of these resonators may be higher as well, which implies that the figures of merit for the devices presented here should be taken as a lower bound. The Si-fin geometry  allows for convenient and accurate capacitor definition as the capacitance is determined by the metallized length along the fins which is set using optical lithography in a subtractive process. The process presented may be used for applications that require a scalable process for compact, low-loss capacitors.

\section{Competing Interests and Data Availability}
All authors declare no financial or non-financial competing interests. The datasets used and/or analysed during the current study available from the corresponding author on reasonable request. This work is a contribution of the U.S. government and is not subject to U.S. copyright.

\begin{acknowledgments}
This work was supported by the U.S. Army Research Office (grant number W911NF-22-1-0052). We wish to acknowledge Jim Beall for assistance with LPCVD silicon nitride growth and useful discussions as well as Dave Pappas. We also acknowledge Pete Hopkins and Chris Yung for a critical reading of this manuscript.
\end{acknowledgments}

\clearpage
\bibliography{main}

\end{document}